\documentclass[review,12pt,authoryear]{elsarticle}

\usepackage{amssymb}
\usepackage{amsthm}
\usepackage{framed} 
\usepackage{amsmath,color}
\usepackage{mathrsfs}
\usepackage{graphicx}
\usepackage{epstopdf}
\usepackage{float}
\usepackage{caption}
\usepackage{subcaption}
\usepackage{bm}
\usepackage{bbm}
\usepackage{mathrsfs}
\usepackage{cleveref}
\usepackage{soul}
\usepackage{accents}
\usepackage{color,soul} 
\usepackage{color} 
\usepackage{nomencl} 
\biboptions{sort&compress}
\soulregister\citep7 
\soulregister\citet7 
\soulregister\citealp7 
\newsavebox{\measurebox} 
\usepackage{titlesec} 
\usepackage{tabu} 
\usepackage{longtable}
\usepackage{multirow} 
\usepackage[nomarkers,figuresonly]{endfloat} 

\journal{European Journal of Mechanics - A/Solids}

\makeatletter
\def\@author#1{\g@addto@macro\elsauthors{\normalsize%
    \def\baselinestretch{1}%
    \upshape\authorsep#1\unskip\textsuperscript{%
      \ifx\@fnmark\@empty\else\unskip\sep\@fnmark\let\sep=,\fi
      \ifx\@corref\@empty\else\unskip\sep\@corref\let\sep=,\fi
      }%
    \def\authorsep{\unskip,\space}%
    \global\let\@fnmark\@empty
    \global\let\@corref\@empty  
    \global\let\sep\@empty}%
    \@eadauthor={#1}
}
\makeatother

\titleformat{\paragraph}
{\normalfont\normalsize\itshape}{\theparagraph}{1em}{}
\titlespacing*{\paragraph}
{0pt}{3.25ex plus 1ex minus .2ex}{1.5ex plus .2ex}

\begin{document}

\begin{frontmatter}



\title{Mode I crack tip fields: strain gradient plasticity theory versus J2 flow theory\footnote{This article was presented at the IUTAM Symposium on Size-Effects in Microstructure and Damage Evolution at Technical University of Denmark, 2018}}


\author{Emilio Mart\'{\i}nez-Pa\~neda\corref{cor1}\fnref{Cam}}
\ead{mail@empaneda.com}

\author{Norman A. Fleck\fnref{Cam}}

\address[Cam]{Department of Engineering, Cambridge University, CB2 1PZ Cambridge, UK}

\cortext[cor1]{Corresponding author.}

\begin{abstract}
The mode I crack tip asymptotic response of a solid characterised by strain gradient plasticity is investigated. It is found that elastic strains dominate plastic strains near the crack tip, and thus the Cauchy stress and the strain state are given asymptotically by the elastic $K$-field. This crack tip elastic zone is embedded within an annular elasto-plastic zone. This feature is predicted by both a crack tip asymptotic analysis and a finite element computation. When small scale yielding applies, three distinct regimes exist: an outer elastic $K$ field, an intermediate elasto-plastic field, and an inner elastic $K$ field. The inner elastic core significantly influences the crack opening profile. Crack tip plasticity is suppressed when the material length scale $\ell$ of the gradient theory is on the order of the plastic zone size estimation, as dictated by the remote stress intensity factor. A generalized $J$-integral for strain gradient plasticity is stated and used to characterise the asymptotic response ahead of a short crack. Finite element analysis of a cracked three point bend specimen reveals that the crack tip elastic zone persists in the presence of bulk plasticity and an outer $J$-field.
\end{abstract}


\end{frontmatter}



\section{Introduction}
\label{Sec:Introduction}

Strain gradient plasticity is increasingly used in fracture analyses to predict the stress elevation that accompanies gradients of plastic strain, see, for example, (\citealp{Wei1997}; \citealp{Jiang2001}; \citealp{Komaragiri2008}; \citealp{Nielsen2012}; \citealp{CM2017}) and references therein. Gradients of plastic strain are associated with lattice curvature and geometrically necessary dislocations \citep{Ashby1970}, and the increased dislocation density promotes strengthening. Flow stress elevation in the presence of plastic strain gradients has been observed in a wide range of mechanical tests on micro-sized specimens. Representative examples are indentation (\citealp{Poole1996}; \citealp{Nix1998}), torsion \citep{Fleck1994}, and bending \citep{Stolken1998}. These experiments typically predict a 3-fold increase in the effective flow stress by reducing the size of the specimen (\emph{smaller is stronger}). Isotropic, strain gradient plasticity theories have been developed to capture this size effect. The pivotal step in constructing these phenomenological models is to write the plastic work increment in terms of both the plastic strain and plastic strain gradient, thereby introducing a length scale in the material description
(\citealp{Aifantis1984}; \citealp{Gao1999}; \citealp{Fleck2001}; \citealp{Gurtin2005}). Work-conjugate stress quantities for plastic strain and plastic strain gradient follow immediately. \\

The crack tip stress elevation, as predicted by strain gradient plasticity theory relative to conventional plasticity theory, plays a fundamental role in the modelling of numerous damage mechanisms \citep{IJSS2015,IJP2016}. Examples include fatigue \citep{Brinckmann2008,Pribe2019}, notch mechanics \citep{TAFM2017}, microvoid cracking \citep{Tvergaard2008}, and hydrogen embrittlement \citep{AM2016,IJHE2016}.\\

In the present study, we examine the mode I crack tip field according to strain gradient plasticity theory (\citealp{Gudmundson2004}; \citealp{Fleck2009}). Previous crack tip asymptotic studies considered earlier gradient plasticity classes, such as couple-stress theories without stretch gradients \citep{Xia1996,Huang1997} or models involving the gradients of elastic strains \citep{Chen1999}. For such theories, plastic strains dominate elastic strains near the crack tip and the asymptotic nature of the crack tip field can be obtained by neglecting elasticity. This is analogous to the HRR \citep{Hutchinson1968,Rice1968} analysis for a conventional elasto-plastic solid.\\

We shall show in the present study that the crack tip field for Gudmundson-type strain gradient theories is of a different nature, such that the asymptotic crack tip field comprises both elastic and plastic straining, and it is not possible to simplify the crack tip asymptotic state by neglecting elastic strains. Instead, the elastic strain $\varepsilon_{ij}^e$ scales as $r^{-1/2}$ with distance $r$ from the crack tip, whereas the plastic strain tensor $\underset{\sim}{\varepsilon}^p \left( \underline{x} \right)$ is of the form
\begin{equation}
\underset{\sim}{\varepsilon}^p \left( \underline{x} \right) = \underset{\sim}{\varepsilon}^p \left( 0 \right) + \underset{\sim}{\tilde{\varepsilon}}^p \left( \underline{x} \right) + \cdots
\end{equation}

The leading order term $\underset{\sim}{\varepsilon}^p \left( 0 \right)$ has a finite value independent of $\underline{x}$. The next term in the series, $\underset{\sim}{\tilde{\varepsilon}}^p \left( \underline{x} \right)$, scales as $r^{3/2}$ where $r = |x|$ is the polar coordinate from the crack tip, and $\underset{\sim}{\tilde{\varepsilon}}^p \left( \underline{x} \right)$ also depends upon the polar coordinate $\theta$. Thus, we can write $\varepsilon^p_{ij}$ in polar coordinates as,
\begin{equation}\label{eq:eprrepthetatheta}
- \varepsilon^p_{rr} = \varepsilon_{\theta \theta}^p= A \cos \left( 2 \theta \right) + r^{3/2} f \left( \theta \right) + \cdots 
\end{equation}
\begin{equation}\label{eq:eprtheta}
\varepsilon_{r \theta}^p = A \sin \left( 2 \theta \right) + r^{3/2} g \left( \theta \right) + \cdots
\end{equation}

Later, in the paper, we shall obtain explicit expressions for the angular functions $f \left( \theta \right)$ and $g \left( \theta \right)$. Thus, the elastic strain is more singular than the plastic strain and the Cauchy stress $\sigma_{ij} \left( r, \theta \right)$ is given by the usual elastic $K$-field in the vicinity of the crack tip.\\

The following simple argument supports the finding that the crack tip is surrounded by an elastic $K$-field in an elastic-plastic strain gradient solid. Introduce a generalized effective plastic strain $\tilde{E}^p$ such that
\begin{equation}\label{Eq:EpGudDef}
\left(\tilde{E}^p \right)^2 = \frac{2}{3} \varepsilon_{ij}^p   \varepsilon_{ij}^p + \ell^2  \varepsilon_{ij,k}^p \varepsilon_{ij,k}^p
\end{equation}

\noindent in terms of a material length scale $\ell$; the comma subscript $()_{,k}$  denotes spatial differentiation with respect to the coordinate $x_k$ in the usual manner. Consider the case of a deformation theory solid, and assume that the plastic strain energy density $w^p$ scales as ${\tilde{E}^p}^{(N+1)}$ in terms of a strain hardening exponent $N$ (where $0 \leq N \leq 1$). We proceed to show that the elastic strain must dominate the plastic strain. To do so, we shall explore the consequences of assuming that the plastic strain dominates the elastic strain near the crack tip. Then, $w^p$ must scale as $J/r$ in order for the energy release rate for crack advance to be finite at the crack tip. Consequently, $\tilde{E}^p$ and $\ell \varepsilon_{ij,k}^p$ scale as $r^{-1/(N+1)}$ and $\varepsilon_{ij}^p$ scales as $r^{N/(N+1)}$. We conclude that $\varepsilon_{ij}^p$ tends to zero as the crack tip is approached. If the elastic strain is dominated by the plastic strain then this implies that $\varepsilon_{ij}^e$ tends to zero at a faster rate than $r^{N/(N+1)}$, and the crack tip will have a strain and a stress concentration of zero. This is implausible on physical grounds. We conclude that the elastic strain field must dominate the plastic strain field at the crack tip, and the Cauchy stress and elastic strain are given by the usual elastic $K$-field.

\section{Strain Gradient Plasticity}
\label{Sec:Theory}

We idealise strain gradient effects by means of the \citet{Gudmundson2004} higher order gradient plasticity model, see also \citet{Fleck2009}. A brief summary of the constitutive and field equations for a flow theory version of strain gradient plasticity is now presented.

\subsection{Variational principles and balance equations}

The primal kinematic variables are the velocity $\dot{u}_i$ and the plastic strain rate $\dot{\varepsilon}_{ij}^p$. Upon adopting a small strain formulation, the total strain rate reads
\begin{equation}
\dot{\varepsilon}_{ij}= \frac{1}{2} \left(\dot{u}_{i,j} + \dot{u}_{j,i} \right)
\end{equation}

\noindent and is decomposed additively into elastic and plastic parts,
\begin{equation}
\dot{\varepsilon}_{ij}=\dot{\varepsilon}_{ij}^e + \dot{\varepsilon}_{ij}^p
\end{equation}

Write the internal work within a volume $V$ as
\begin{equation}\label{eq:PVW1Gud}
\delta W = \int_V \Big( \sigma_{ij} \delta \varepsilon_{ij}^e + q_{ij} \delta \varepsilon_{ij}^p + \tau_{ijk} \delta \varepsilon_{ij,k}^p \Big) dV
\end{equation}

\noindent where $\sigma_{ij}$ denotes the Cauchy stress, $q_{ij}$ the so-called micro-stress tensor (work-conjugate to the plastic strain $\varepsilon_{ij}^p$) and $\tau_{ijk}$ is the higher order stress tensor (work-conjugate to the plastic strain gradient $\varepsilon_{ij,k}^p$). The volume $V$ is contained within a surface $S$ of unit outward normal $n_i$. Now make use of Gauss' divergence theorem to re-express $\delta W$ as the external work on the surface $S$,
\begin{equation}\label{eq:PVW2Gud}
\delta W = \int_S \left( \sigma_{ij} n_j \delta u_i + \tau_{ijk} n_k \delta \varepsilon_{ij}^p \right) dS
\end{equation}

\noindent to obtain the following equilibrium equations within $V$:
\begin{align}
& \sigma_{ij,j}=0 \nonumber \\ 
& s_{ij} = q_{ij} - \tau_{ijk,k} \label{eq:EQ2Gud}
\end{align}

\noindent Here, $s_{ij}$ is the deviatoric part of the Cauchy stress such that $s_{ij}=\sigma_{ij}-\delta_{ij} \sigma_{kk}/3$. Equations (\ref{eq:PVW1Gud}) and (\ref{eq:PVW2Gud}) constitute the Principle of Virtual Work,
\begin{equation}
\int_V \Big( \sigma_{ij} \delta \varepsilon_{ij}^e + q_{ij} \delta \varepsilon_{ij}^p + \tau_{ijk} \delta \varepsilon_{ij,k}^p \Big) dV = \int_S \left( T_i \delta u_i + t_{ij} \delta \varepsilon_{ij}^p \right) dS
\end{equation}

\noindent where $T_i = \sigma_{ij} n_j $ and $t_{ij} = \tau_{ijk} n_k$ denote the conventional and higher order tractions, respectively.

\subsection{Constitutive laws}

The elastic strain $\varepsilon_{ij}^e$ gives rise to an elastic strain energy density,
\begin{equation}\label{eq:FreeEnergy}
w^e=\frac{1}{2} \varepsilon_{ij}^e C_{ijkl} \varepsilon_{kl}^e
\end{equation}

\noindent where $C_{ijkl}=C_{klij}$ is the isotropic elastic stiffness tensor, given in terms of Young's modulus $E$ and Poisson's ratio $\nu$. We identify the elastic work increment $\sigma_{ij} \delta \varepsilon_{ij}^e$ with $\delta w^e$ such that
\begin{equation}\label{eq:HookeLaw}
\sigma_{ij}=\frac{\partial w^e}{\partial \varepsilon_{ij}^e}=C_{ijkl}\varepsilon_{kl}^e
\end{equation}

The stresses $\left( q_{ij}, \, \tau_{ijk} \right)$ are taken to be dissipative in nature and we assume that the plastic work rate $\dot{w}^p$ reads,
\begin{equation}\label{Eq:PlasticWork}
q_{ij} \delta \dot{\varepsilon}_{ij}^p + \tau_{ijk} \delta \dot{\varepsilon}_{ij,k}^p = \delta \dot{w}^p
\end{equation}

\noindent where $\dot{w}^p (\dot{E}^p)$ is given in terms of a combined effective plastic strain rate,
\begin{equation}\label{Eq:EpGud}
\dot{E}^p=\left( \frac{2}{3} \dot{\varepsilon}^p_{ij} \dot{\varepsilon}^p_{ij} + \ell^2 \dot{\varepsilon}^p_{ij,k} \dot{\varepsilon}^p_{ij,k} \right)^{1/2}
\end{equation}

\noindent thereby introducing a material length scale $\ell$. The use of (\ref{Eq:PlasticWork}) implies immediately that
\begin{equation}
q_{ij} = \frac{\partial \dot{w}^p}{\partial \dot{\varepsilon}_{ij}^p} = \frac{\partial \dot{w}^p}{\partial \dot{E}^p} \frac{\partial \dot{E}^p}{\partial \dot{\varepsilon}_{ij}^p}
\end{equation}

\noindent and
\begin{equation}
\tau_{ijk} = \frac{\partial \dot{w}^p}{\partial \dot{\varepsilon}_{ij,k}^p} = \frac{\partial \dot{w}^p}{\partial \dot{E}^p} \frac{\partial \dot{E}^p}{\partial \dot{\varepsilon}_{ij,k}^p}
\end{equation}

Upon introducing an overall effective stress $\Sigma = \partial \dot{w}^p / \partial \dot{E}^p$, these expressions reduce to
\begin{equation}\label{Eq:qGud}
q_{ij}= \frac{2}{3} \frac{\Sigma}{\dot{E}^p}\dot{\varepsilon}^p_{ij} \,\,\,\,\,\, \textnormal{and} \,\,\,\,\,\, \tau_{ijk}=\frac{\Sigma}{\dot{E}^p} \ell^2 \dot{\varepsilon}^p_{ij,k}
\end{equation}

Note that $\Sigma$ is work conjugate to $\dot{E}^p$, such that it satisfies
\begin{equation}
\Sigma \dot{E}^p = q_{ij} \dot{\varepsilon}_{ij}^p + \tau_{ijk} \dot{\varepsilon}_{ij,k}^p
\end{equation}

\noindent and, upon making use of (\ref{Eq:EpGud}) and (\ref{Eq:qGud}) we obtain the relation
\begin{equation}\label{Eq:Sigma}
\Sigma=\left( \frac{3}{2} q_{ij} q_{ij} + \ell^{-2} \tau_{ijk} \tau_{ijk} \right)^{1/2}
\end{equation}

\section{Asymptotic analysis of crack tip fields}
\label{Sec:AResults}

\subsection{Deformation theory solid}

We begin our study by conducting an asymptotic analysis of the stress and strain state at the crack tip. As already discussed in the introduction, consider a deformation theory solid such that the strain energy density $w \left(\varepsilon_{ij}^e,\, \varepsilon_{ij}^p, \, \varepsilon_{ij,k}^p  \right)$ is decomposed into an elastic part $w^e$ and a plastic part $w^p$,
\begin{equation}\label{Eq:W}
w \left( \varepsilon_{ij}^e, \, \varepsilon_{ij}^p, \, \varepsilon_{ij,k}^p \right) =  w^e   \left( \varepsilon_{ij}^e \right) + w^p \left( \varepsilon_{ij}^p, \, \varepsilon_{ij,k}^p \right)
\end{equation}

The elastic contribution is stated explicitly by (\ref{eq:FreeEnergy}). For the deformation theory solid the effective strain quantity $\tilde{E}^p$ has already been introduced by (\ref{Eq:EpGudDef}). The dissipation potential $w^p$ is taken to be a power law function of $\tilde{E}^p$
\begin{equation}\label{Eq:phi}
w^p \left( \tilde{E}^p \right) = \frac{\sigma_Y \varepsilon_Y}{N+1} \left( \frac{\tilde{E}^p}{\varepsilon_Y} \right)^{N+1}
\end{equation}

\noindent in terms of a reference value of strength $\sigma_Y$, yield strain $\varepsilon_Y=\sigma_Y/E$ and hardening index $N$ (where $0 \leq N \leq 1$). Upon writing the dissipation increment $\delta w^p$ as 
\begin{equation}\label{Eq:phiInc}
\delta w^p = q_{ij} \delta \varepsilon_{ij}^p + \tau_{ijk} \delta \varepsilon_{ij,k}^p
\end{equation}

\noindent and upon introducing the notation $\Sigma = \partial w^p / \partial \tilde{E}^p$, we have
\begin{equation}\label{Eq:qDef}
q_{ij} = \frac{\partial w^p}{\partial \varepsilon_{ij}^p} = \Sigma \frac{\partial \tilde{E}^p}{\partial \varepsilon_{ij}^p} = \frac{2}{3} \frac{\Sigma}{\tilde{E}^p} \varepsilon_{ij}^p
\end{equation}
\begin{equation}\label{Eq:tauDef}
\tau_{ijk} = \frac{\partial w^p}{\partial \varepsilon_{ij,k}^p} = \Sigma \frac{\partial \tilde{E}^p}{\partial \varepsilon_{ij,k}^p} = \ell^2 \frac{\Sigma}{\tilde{E}^p} \varepsilon_{ij,k}^p
\end{equation}

We note in passing that substitution of (\ref{Eq:qDef})-(\ref{Eq:tauDef}) into (\ref{Eq:EpGudDef}) recovers (\ref{Eq:Sigma}), and the relation between $\Sigma$ and $\tilde{E}^p$ is of power law type, such that
\begin{equation}
\Sigma =\frac{\partial w^p}{\partial \tilde{E}^p} = \sigma_Y \left( \frac{\tilde{E}^p}{\varepsilon_Y} \right)^N
\end{equation}

\noindent via (\ref{Eq:phiInc}).

\subsection{Energy boundness analysis}

We proceed to obtain the asymptotic nature of $\varepsilon_{ij}^p \left( r , \theta \right)$. The finite element solutions presented later in the study consistently reveal that the deviatoric Cauchy stress $s_{ij}$ scales as $r^{-1/2}$. We shall adopt this scaling law for $s_{ij}$ and explore its ramifications. First, note from (\ref{Eq:qDef}) and (\ref{Eq:tauDef}) that $\tau_{ijk}$, and consequently $\tau_{ijk,k}$, are more singular in $r$ than $q_{ij}$, as the crack tip is approached. Then, the equilibrium relation (\ref{eq:EQ2Gud})b demands that
\begin{equation}\label{eq:EQ2GudAsymp}
s_{ij} \sim - \tau_{ijk,k}
\end{equation}

\noindent to leading order in $r$, and consequently $\tau_{ijk}$ is of order $r^{1/2}$. This imposes a severe restriction on the form of $\varepsilon_{ij}^p \left( r , \theta \right)$. Assume the separation of variables form for $\varepsilon_{ij}^p$ in terms of its Cartesian components
\begin{equation}\label{Eq:PlasticStrainExpansion}
\varepsilon_{ij}^p = A_{ij} + r^{\alpha} B_{ij} \left( \theta \right) + \cdots
\end{equation}

\noindent where $A_{ij}$ is taken to be independent of $\theta$ and the index $\alpha>0$ remains to be found. First we show that this form satisfies the field equations, and second we justify this choice. Accordingly, the plastic strain gradient reads
\begin{equation}\label{Eq:GradientExpansion}
\varepsilon_{ij,k}^p = \alpha r^{\alpha -1} \bar{B}_{ijk} \left( \theta \right) + \cdots
\end{equation}

\noindent where $\bar{B}_{ijk}$ can be expressed in terms of $B_{ij} \left( \theta \right)$ and its derivatives with respect to $\theta$. Substitution of (\ref{Eq:PlasticStrainExpansion}) and (\ref{Eq:GradientExpansion}) into (\ref{Eq:EpGudDef}) gives
\begin{equation}\label{Eq:EpAsymp1}
\left(\tilde{E}^p \right)^2 = \frac{4}{3} A_{ij} A_{ij} + \cdots
\end{equation}

\noindent along with
\begin{equation}
\frac{\Sigma}{E_p} = \sigma_Y \left( \frac{2}{\sqrt{3} \varepsilon_Y} \right)^{N-1} \left( A_{ij} A_{ij} \right)^{\frac{N-1}{2}}
\end{equation}

Consequently, (\ref{eq:EQ2GudAsymp}), (\ref{Eq:tauDef}) and (\ref{Eq:GradientExpansion}) give
\begin{equation}
s_{ij} = -\ell^2 \sigma_Y \left( \frac{2}{\sqrt{3} \varepsilon_Y}\right)^{N-1} \left( A_{pq} A_{pq} \right)^{\frac{N-1}{2}} \left( r^{\alpha -1} \bar{B}_{ijk} \left( \theta \right) \right)_{,k}
\end{equation}

Upon recalling that $s_{ij}$ scales as $r^{-1/2}$ the above equation implies that $\alpha=3/2$ for consistency. The above solution reveals that the elastic strain energy density $w^E$ scales as $r^{-1}$ while the plastic strain energy density scales as $r^0$, upon recalling (\ref{Eq:phi}) and (\ref{Eq:EpAsymp1}). Now recall that we require $w \sim J/r$ in order for $w=w^e+w^p$ to give a finite energy release rate $J$ at the crack tip. This is achieved by the elastic field whereas the plastic field is not sufficiently singular in $r$ to give any contribution to the energy release rate. Alternative assumptions can be made for the series expansion of $\varepsilon_{ij}^p$ in preference to (\ref{Eq:PlasticStrainExpansion}). However, these do not give rise to an equilibrium solution (i.e., Eq. (\ref{eq:EQ2GudAsymp}) is not satisfied) or they give solutions that are less singular than that of (\ref{Eq:EpAsymp1}). For example, if we assume that the Cartesian components $A_{ij}$ are a function of $\theta$ we find that $\varepsilon_{ij,k}^p$ scales as $r^{-1}$ and $\tau_{ijk,k}$ scales as $r^{-(N+1)}$, and the equilibrium relation (\ref{Eq:EpAsymp1}) for $s_{ij}$ is violated. Alternatively, if we take $A_{ij} =0$ then an equilibrium solution for $s_{ij}$ is obtained provided we take $\alpha=N/(N+1)$. This leads to a higher order term in the series expansion of $\varepsilon_{ij}^p$ than that given by the first two terms of (\ref{Eq:EpAsymp1}). Finally, what is the implication of assuming that $\alpha <0$ in our asymptotic expression (\ref{Eq:EpAsymp1})? If we were to assume $\alpha<0$, then the leading order term becomes $r^\alpha B_{ij} \left( \theta \right)$. Asymptotic matching of both sides of the equilibrium relation (\ref{eq:EQ2GudAsymp}) again results in $\alpha=N/(N+1)$, which is inconsistent with the initial assumption that $\alpha<0$.\\

In summary, the plastic strain field $\varepsilon_{ij}^p \left( r, \theta \right)$ is of the asymptotic form (\ref{Eq:PlasticStrainExpansion}) with $\alpha=3/2$, and the crack tip field for Cauchy stress $\sigma_{ij} \left( r, \theta \right)$ and elastic strain $\varepsilon^e_{ij} \left( r, \theta \right)$ is given by the usual $K$-field for a mode I crack.

\subsection{Asymptotic crack tip fields}

Assume that the leading order terms in $\varepsilon_{ij}^p$, in polar coordinates, are of the form (\ref{eq:eprrepthetatheta})-(\ref{eq:eprtheta}). This choice is consistent with the nature of the symmetry of the solution of a mode I crack tip problem; $\varepsilon_{\theta \theta}^p$ and $\varepsilon_{rr}^p$ are even in $\theta$ and give rise to $\varepsilon_{yy}^p=A=-\varepsilon_{xx}^p$, $\varepsilon^p_{xy}=0$ in Cartesian coordinates. The components of the plastic strain gradient and the Laplacian of the plastic strain read
\begin{equation}
\varepsilon^p_{rr,r} = \frac{\partial \varepsilon^p_{rr}}{\partial r} = - \frac{3}{2} r^{1/2} f \left( \theta \right)
\end{equation}
\begin{equation}
\varepsilon^p_{rr,\theta} = \frac{1}{r} \left(  \frac{\partial \varepsilon^p_{rr}}{\partial \theta} - 2  \varepsilon^p_{r \theta} \right)  = -  r^{1/2} \left[ 3 f' \left( \theta \right) - 2 g \left( \theta \right) \right]
\end{equation}
\begin{equation}
\varepsilon^p_{r \theta, r} =  \frac{\partial \varepsilon^p_{r\theta}}{\partial r} =  \frac{3}{2} r^{1/2} g \left( \theta \right)
\end{equation}
\begin{equation}
\varepsilon^p_{r \theta, \theta} = \frac{1}{r} \left( \frac{\partial \varepsilon^p_{r \theta}}{\partial \theta}+ 2\varepsilon^p_{rr} \right) =  r^{1/2} \left[ g ' \left( \theta \right) - f \left( \theta \right) \right] 
\end{equation}

\noindent and,

\begin{align}\label{eq:eprr,kk}
\varepsilon^p_{rr,kk} &= \frac{\partial^2 \varepsilon^p_{rr}}{\partial r^2} + \frac{1}{r} \frac{\partial \varepsilon_{rr}^p}{\partial r} + \frac{1}{r^2}\frac{\partial^2 \varepsilon_{rr}^p}{\partial \theta^2} -  \frac{2}{r^2}\left( \varepsilon^p_{r r} - \varepsilon^p_{\theta \theta} + 2 \frac{\partial \varepsilon^p_{r \theta}}{\partial \theta}\right) \\ \nonumber 
&= r^{-1/2} \left[ \frac{7}{4} f \left( \theta \right) - f '' \left( \theta \right) +  4 g' \left( \theta \right) \right] 
\end{align}
\begin{align}\label{eq:eprtheta,kk}
\varepsilon^p_{r \theta,kk} & = \frac{\partial^2 \varepsilon^p_{r\theta}}{\partial r^2} +  \frac{1}{r} \frac{\partial \varepsilon_{r\theta}^p}{\partial r} + \frac{1}{r^2} \frac{\partial^2 \varepsilon^p_{r \theta}}{\partial \theta^2} + \frac{2}{r^2} \left(\frac{\partial \varepsilon^p_{rr}}{\partial \theta} -\frac{\partial \varepsilon^p_{\theta \theta}}{\partial \theta} - 2 \varepsilon^p_{r \theta} \right) \\ \nonumber 
& = r^{-1/2} \left[ - \frac{7}{4}  g \left( \theta \right) +  g'' \left( \theta \right) - 4  f' \left( \theta \right)\right]
\end{align}

Now make use of the higher order equilibrium equation (\ref{eq:EQ2Gud})b, which asymptotically implies $s_{ij} \approx - \tau_{ijk,k}$. Note that, as $r \to 0$,  $\tilde{E}^p$ is of leading order $ 2A/ \sqrt{3} $ and can therefore be treated as a constant. As argued above and demonstrated numerically below, the Cauchy stress is characterized by an inner elastic $K$-field. Consequently, we make use of the \citet{Williams1957} solution to write
\begin{equation}\label{eq:srrDef}
s_{rr} = \frac{K_I}{4 \sqrt{2 \pi r}} \left[ \cos \left( \frac{\theta}{2} \right) - \cos \left( \frac{3 \theta}{2} \right) \right]
\end{equation}
\begin{equation}\label{eq:srthetaDef}
s_{r \theta}= \frac{K_I}{4 \sqrt{2 \pi r}} \left[ \sin \left( \frac{\theta}{2} \right) + \sin \left( \frac{3 \theta}{2} \right) \right]
\end{equation}

\noindent where $K_I$ is the mode I stress intensity factor. The higher order equilibrium follows by suitable substitution of (\ref{eq:eprr,kk})-(\ref{eq:srthetaDef}) into (\ref{Eq:tauDef}) and (\ref{eq:EQ2Gud})b, to give
\begin{equation}
\frac{K_I \tilde{E}^p}{4 \Sigma \sqrt{2 \pi}} \left[ \cos \left( \frac{\theta}{2} \right) - \cos \left( \frac{3 \theta}{2} \right) \right] = - \frac{7}{4} f \left(\theta \right)  + \frac{\partial^2 f \left( \theta \right)}{\partial \theta^2} + 4 \frac{\partial g \left( \theta \right)}{\partial \theta}
\end{equation}
\begin{equation}
\frac{K_I \tilde{E}^p}{4 \Sigma \sqrt{2 \pi}}\left[ \sin \left( \frac{\theta}{2} \right) + \sin \left( \frac{3 \theta}{2} \right) \right]  = \frac{7}{4} g \left(\theta \right) -\frac{\partial^2 g \left( \theta \right)}{\partial \theta^2} + 4 \frac{\partial f \left( \theta \right)}{\partial \theta}
\end{equation}

In addition, the symmetry condition ahead of a mode I crack tip demands that $f \left( \theta \right)$ is an even function of $\theta$. Thus, the solution to the system of differential equations is given by
\begin{equation}
f \left( \theta \right) = - \frac{K_I \tilde{E}^p}{16 \Sigma \sqrt{2 \pi}} \left[2 \cos \left( \frac{3 \theta}{2} \right)+ \cos \left( \frac{\theta}{2} \right)  + a_1 \cos \left( \frac{\theta}{2}\right) + a_3 \cos \left( \frac{7 \theta}{2} \right) \right] 
\end{equation}
\begin{equation}
g \left( \theta \right) = - \frac{K_I \tilde{E}^p}{16 \Sigma \sqrt{2 \pi}} \left[2 \sin \left( \frac{3 \theta}{2} \right) - \sin \left( \frac{\theta}{2} \right) + a_1 \sin \left( \frac{\theta}{2}\right) + a_3 \sin \left( \frac{7 \theta}{2} \right) \right] 
\end{equation}

Now make use of the traction-free boundary conditions $t_{rr}=t_{\theta \theta}=0$ along the crack flanks ($\theta = \pi$) to obtain a relation between the constants $a_1$ and $a_3$. Free boundary conditions on the higher order traction $\tau_{\theta \theta \theta}=0$ on $\theta=\pi$ implies that
\begin{equation}
f' \left( \theta \right) + 2 g \left( \theta \right) = 0 \,\,\,\,\,\,\, \text{at} \,\,\,\,\,\,\, \theta=\pi
\end{equation}

\noindent rendering $a_1=7/3-a_3$. Imposition of vanishing higher order traction $\tau_{r \theta \theta}=0$ on $\theta=\pi$ is identically satisfied and provides no useful additional information on $(a_1, a_3)$. It follows that numerical analysis is needed to calibrate $a_1$ and $a_3$ and obtain a full field solution. 

\section{Finite element analysis}
\label{Sec:NumericalResults}

\subsection{Numerical implementation}

We make use of the finite element implementation of \citet{JMPS2019} and employ the viscoplastic potential presented by \citet{Panteghini2016}. The effective stress is related to the gradient-enhanced effective plastic flow rate through a viscoplastic function,
\begin{equation}
\Sigma=\sigma_F \left( E^p \right) V(\dot{E}^p)
\end{equation}

\noindent where the current flow stress $\sigma_F$ depends on the initial yield stress $\sigma_Y$ and on $E^p$ via a hardening law. We adopt the following isotropic hardening law,
\begin{equation}\label{Eq:IsoHard}
\sigma_F=\sigma_Y \left( 1 + \frac{E^p}{\varepsilon_Y} \right)^N
\end{equation}

\noindent and assume that the yield strain is $\varepsilon_Y=\sigma_Y/E=0.003$. The viscoplastic function $V (\dot{E}^p)$ is defined as
\begin{equation}
V (\dot{E}^p) =   \begin{cases} 
   \dot{E}^p/ \left(2 \dot{\varepsilon}_0 \right) & \text{if } \dot{E}^p/\dot{\varepsilon}_0 \leq 1 \\
   1 - \dot{\varepsilon}_0 / \left( 2 \dot{E}^p \right)     & \text{if } \dot{E}^p / \dot{\varepsilon}_0 > 1
  \end{cases}
\end{equation}
\noindent and the rate-independent limit is achieved by choosing a sufficiently small value of the material parameter $\dot{\varepsilon}_0$. The numerical experiments conducted show that the ratio $\dot{E}^p/\dot{\varepsilon}_0$ is sufficiently high that $V ( \dot{E}^p ) \approx 1$ in the vicinity of the crack for all remote $K$ values considered.\\

A mixed finite element scheme is adopted, such that displacements and plastic strains are treated as primary variables, in accordance with the theoretical framework. The non-linear system of equations for a time $t+\Delta t$ is solved iteratively by using the Newton-Raphson method,
\begin{equation}
\begin{bmatrix} \bm{u} \\ \bm{\varepsilon_p} \end{bmatrix}_{t+\Delta t} = \begin{bmatrix} \bm{u} \\ \bm{\varepsilon_p} \end{bmatrix}_{t} - \begin{bmatrix}
  \bm{K}_{u,u} & \bm{K}_{u,\varepsilon^p}\\
  \bm{K}_{\varepsilon^p,u} & \bm{K}_{\varepsilon^p,\varepsilon^p}
 \end{bmatrix}_t ^{-1}\begin{bmatrix} \bm{R}_u \\ \bm{R}_{\varepsilon^p} \end{bmatrix}_t
\end{equation}

\noindent where the residuals comprise the out-of-balance forces,
\begin{equation}
\bm{R}_u^n=\int_{V} \sigma_{ij} B_{ij}^n \, dV - \int_{S} T_i N_i^n \, dS
\end{equation}
\begin{equation}
\bm{R}_{\varepsilon^p}^n=\int_{V} \left[(q_{ij}-s_{ij})  M_{ij}^n  + \tau_{ijk} M_{ij,k}^n \right] dV - \int_{S} t_{ij} M_{ij,k}^n \, dS
\end{equation}

\noindent Here, $B_{ij}$ denotes the strain-displacement matrix, and $N_i$ and $M_{ij}$ are the shape functions for the nodal values of displacement and plastic strain components. The components of the consistent stiffness matrix are obtained by differentiating the residuals with respect to the incremental nodal variables. The reader is referred to \citet{JMPS2019} for full details.

\subsection{The small scale yielding solution}
\label{Sec:K}

We make use of the so-called boundary layer formulation to prescribe an outer $K$-field. Consider a crack with tip at the origin and with the crack plane along the negative axis of the Cartesian reference frame $(x,y)$. A remote $K$ field is imposed by prescribing the nodal displacements in the outer periphery of the mesh as,
\begin{equation}
u_i = \frac{K}{E} r^{1/2} f_i \left( \theta, \nu \right)
\end{equation}

\noindent where $\nu$ is Poisson's ratio and the functions $f_i \left( \theta, \nu \right)$ are given by
\begin{equation}
f_x = \frac{1+\nu}{\sqrt{2 \pi}} \left(3 - 4 \nu - \cos \theta \right) \, \cos \left(\frac{\theta}{2} \right)
\end{equation}

\noindent and
\begin{equation}
f_y = \frac{1+\nu}{\sqrt{2 \pi}} \left(3 - 4 \nu - \cos \theta \right) \, \sin \left(\frac{\theta}{2} \right)
\end{equation}

Upon exploiting the symmetry about the crack plane, only half of the finite element model is analysed. A mesh sensitivity study reveals that it is adequate to discretise the domain by approximately 5200 plane strain, quadratic, quadrilateral elements.\\

\begin{figure}[H]
\centering
\includegraphics[scale=0.6]{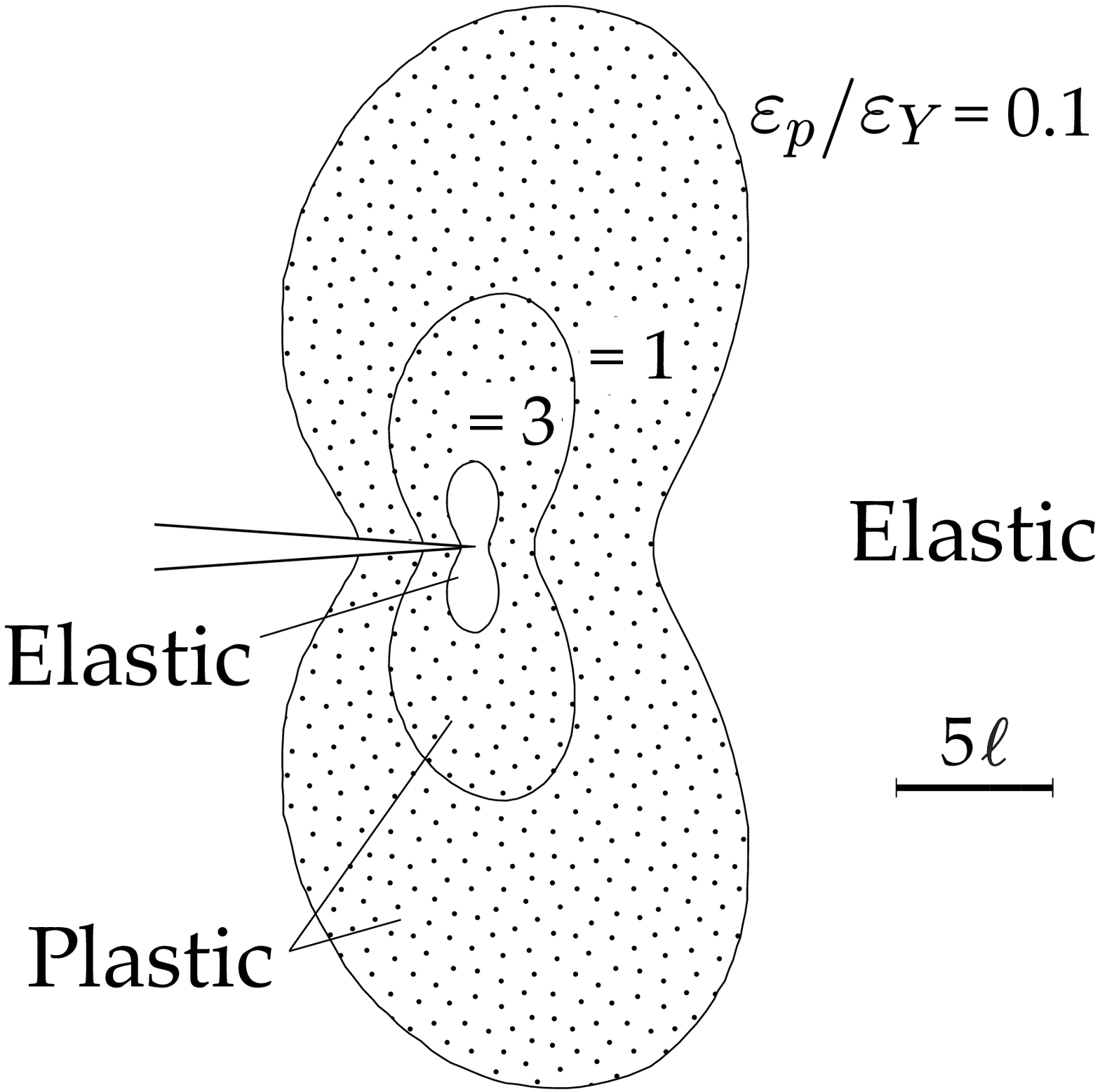}
\caption{Finite element predictions of the different domains surrounding the crack tip in an strain gradient solid at $K=20 \sigma_Y \sqrt{\ell}$. Three regions are identified as a function of the effective von Mises plastic strain $\varepsilon_p$: the outer elastic domain, the plastic zone and the inner elastic core. A scale bar of length $5 \ell$ is included. Material properties: $N=0.1$, $\varepsilon_Y=0.003$, and $\nu=0.3$.}
\label{fig:regimes}
\end{figure}

\begin{figure}[H]
\centering
\includegraphics[scale=0.95]{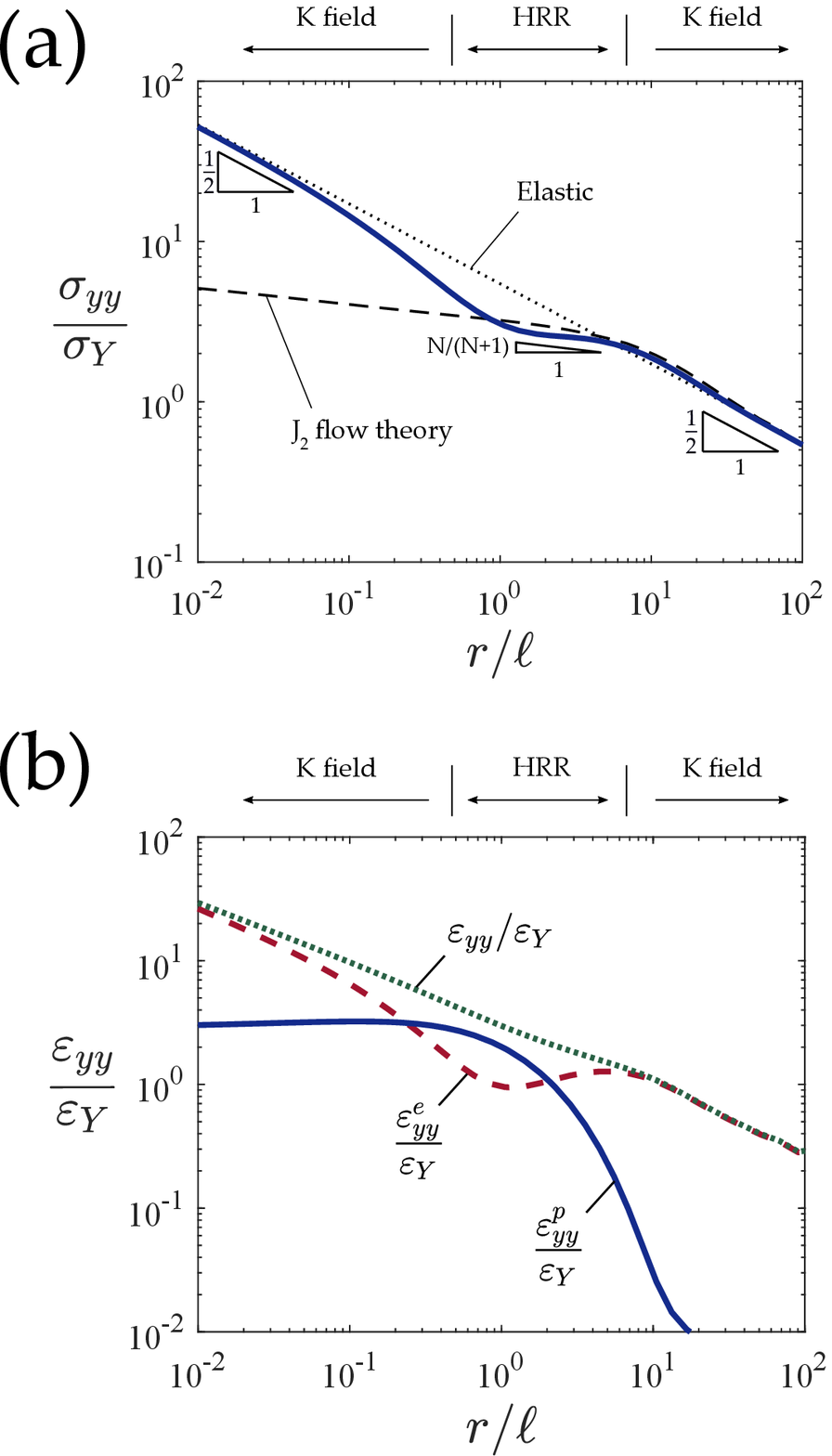}
\caption{Tensile (a) stress and (b) strain ahead of the crack tip for a strain gradient solid at $K=20 \sigma_Y \sqrt{\ell}$. Material properties: $N=0.1$, $\varepsilon_Y=0.003$, and $\nu=0.3$.}
\label{fig:S22vsR}
\end{figure}

A representative small scale yielding solution is now presented in Figs. \ref{fig:regimes} and \ref{fig:S22vsR}, for the choice $K=20 \sigma_Y \sqrt{\ell}$, $N=0.1$, $\varepsilon_Y=0.003$, and $\nu=0.3$. Conventional $J_2$ flow theory implies a plastic zone size $R_0$ of magnitude
\begin{equation}\label{Eq:Irwin}
R_0 =\frac{1}{3 \pi} \left( \frac{K}{\sigma_Y} \right)^2
\end{equation}

\noindent and so the choice $K=20 \sigma_Y \sqrt{\ell}$ implies $R_0=42 \ell$. Consequently, the plastic zone size is much larger than $\ell$ for the strain gradient solid also. The plastic zone is plotted in Fig. \ref{fig:regimes} by showing contours of von Mises plastic strain,
\begin{equation}
\varepsilon_p = \left( \frac{2}{3} \varepsilon_{ij}^p \varepsilon_{ij}^p \right)^{1/2}
\end{equation}

\noindent In broad terms, the outer boundary of the plastic zone is given by the contour $\varepsilon_p/\varepsilon_Y=0.1$. Additional contours for $\varepsilon_p/\varepsilon_Y=1$ and 3 are included. It is found that $\varepsilon_p/\varepsilon_Y$ attains a plateau value slightly greater than 3 within the contour $\varepsilon_p/\varepsilon_Y=3$. Consequently, the stress state within this crack tip zone is elastic in nature. This finding is supported by a plot of tensile stress $\sigma_{yy}$ as a function of $r$ directly ahead of the crack tip ($y=0$), see Fig. \ref{fig:S22vsR}a. The stress component $\sigma_{yy}$ scales as $r^{-1/2}$ for sufficiently small $r$. Likewise, the elastic strain component $\varepsilon_{yy}^e$ scales as $r^{-1/2}$ for $r / \ell < 1$, see Fig. \ref{fig:S22vsR}b. Farther from the crack tip ($1 < r/ \ell <10$) the stress profile $\sigma_{yy}$ varies with $r$ in the manner of the HRR field, $\sigma_{\theta \theta} \sim r^{-N/(N+1)}$. Beyond the plastic zone ($r/\ell>10$) the stress state again converges to the elastic $K$-field and $\sigma_{yy}$ scales as $r^{-1/2}$. Thus, both an outer and an inner $K$ field exist. The distributions of $\varepsilon^e_{yy} (r)$ and $\varepsilon^p_{yy} (r)$ are shown in Fig. \ref{fig:S22vsR}b. Within the elastic zone at the crack tip, and in the outer elastic zone, we have $\varepsilon_{yy}^e >> \varepsilon_{yy}^p$. In contrast, within the annual region of the crack tip plastic zone, the plastic strains dominate and $\varepsilon^p_{yy} > \varepsilon^e_{yy}$.\\

The following $J$-integral argument can be used to show that the magnitude of $K$ for the crack tip elastic zone is identical to that in the outer field. Write the potential energy $P$ of the cracked solid as
\begin{equation}
P (a) = \int_V w \, \text{d}V - \int_{S_T} \left( T_i^\infty u_i + t_{ij}^\infty \varepsilon^p_{ij} \right) \, \text{d}S
\end{equation}

\noindent where $(T_i^\infty, t_{ij}^\infty)$ are the prescribed tractions on a partial boundary $S_T$, with outward normal $n_i$. Define $J$ as the energy release rate per unit crack extension, such that
\begin{equation}
J = -\frac{\partial P}{\partial a}
\end{equation}

\noindent for a body of unit thickness in the $z$ direction. Note that
\begin{equation}
\oint \left( w n_x - \sigma_{ij} n_j u_{i,x} - \tau_{ijk} n_k \varepsilon_{ij,x}^p \right) \, \text{d}S \equiv 0
\end{equation}

\noindent for any closed contour in the solid that excludes the crack tip. Also note that $\sigma_{ij} n_j =0$ and $\tau_{ijk} n_k =0$ on the faces of a traction-free crack. Then, an evaluation of $J$ for a contour $\Gamma$ which encloses the crack tip, starts on the lower crack flank and ends on the upper flank, gives
\begin{equation}
J = \int_\Gamma \left( w n_x - \sigma_{ij} n_j u_{i,x} - \tau_{ijk} n_k \varepsilon^p_{ij,x} \right) \, \text{d} S
\end{equation}

\noindent where the crack lies along the negative $x$-axis. The proof is straightforward and follows that outlined by \citep{Eshelby1956,Rice1968a} for the conventional deformation theory solid, absent strain gradient effects.\\

Now evaluate the contour integral $J$ assuming that the stress state (and associated strain energy density $w$) is given by an elastic $K$-field. Direct evaluation gives the Irwin relation $EJ/(1-\nu^2)=K^2$. Upon performing this integration within the crack tip elastic zone of the strain gradient solid, and repeating the evaluation in the outer $K$-field remote from the crack tip, path independence of $J$ immediately implies that the magnitude of $K$ is the same in the two zones.\\

\subsection{Sensitivity of crack tip fields to strain hardening and material length scale}

\begin{figure}[H]
\centering
\includegraphics[scale=1]{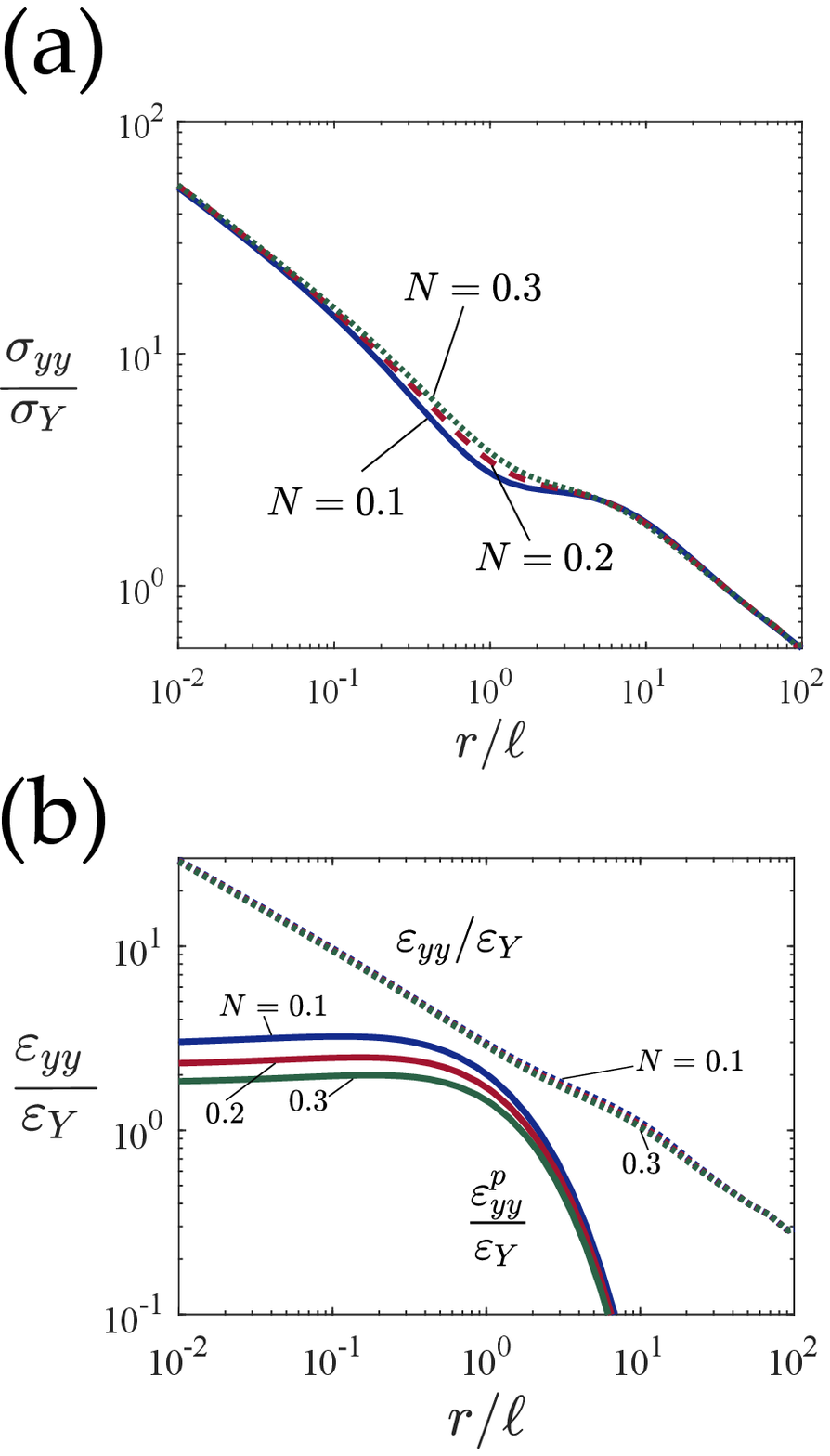}
\caption{Tensile (a) stress and (b) strain ahead of the crack tip for a strain gradient solid with different values of the strain hardening exponent $N$ at $K=20 \sigma_Y \sqrt{\ell}$. Material properties: $N=0.1$, $\varepsilon_Y=0.003$, and $\nu=0.3$.}
\label{fig:N}
\end{figure}

We proceed to examine the influence of the strain hardening exponent $N$ upon the crack tip stress state, see Fig. \ref{fig:N}a. Consistent with the analytical asymptotic analysis of Section \ref{Sec:AResults}, the near-tip asymptotic response is independent of the value of $N$ and the three regimes (outer $K$, elastic-plastic field and inner $K$) can be readily identified for the three values of $N$ considered. The strain state near the crack tip is shown in the form of the components $\varepsilon_{yy}^p$ and $\varepsilon_{yy}$ versus $r/ \ell$ in Fig. \ref{fig:N}b. The asymptotic value of $\varepsilon_{yy}^p (r \to 0)$ increases slightly with decreasing $N$. The zone of almost constant $\varepsilon_{yy}^p$ near the crack tip is of similar size for $N=0.1$, 0.2 and 0.3: the size of the elastic core scales with $\ell$ and is independent of $N$. 

\begin{figure}[H]
\centering
\includegraphics[scale=1.1]{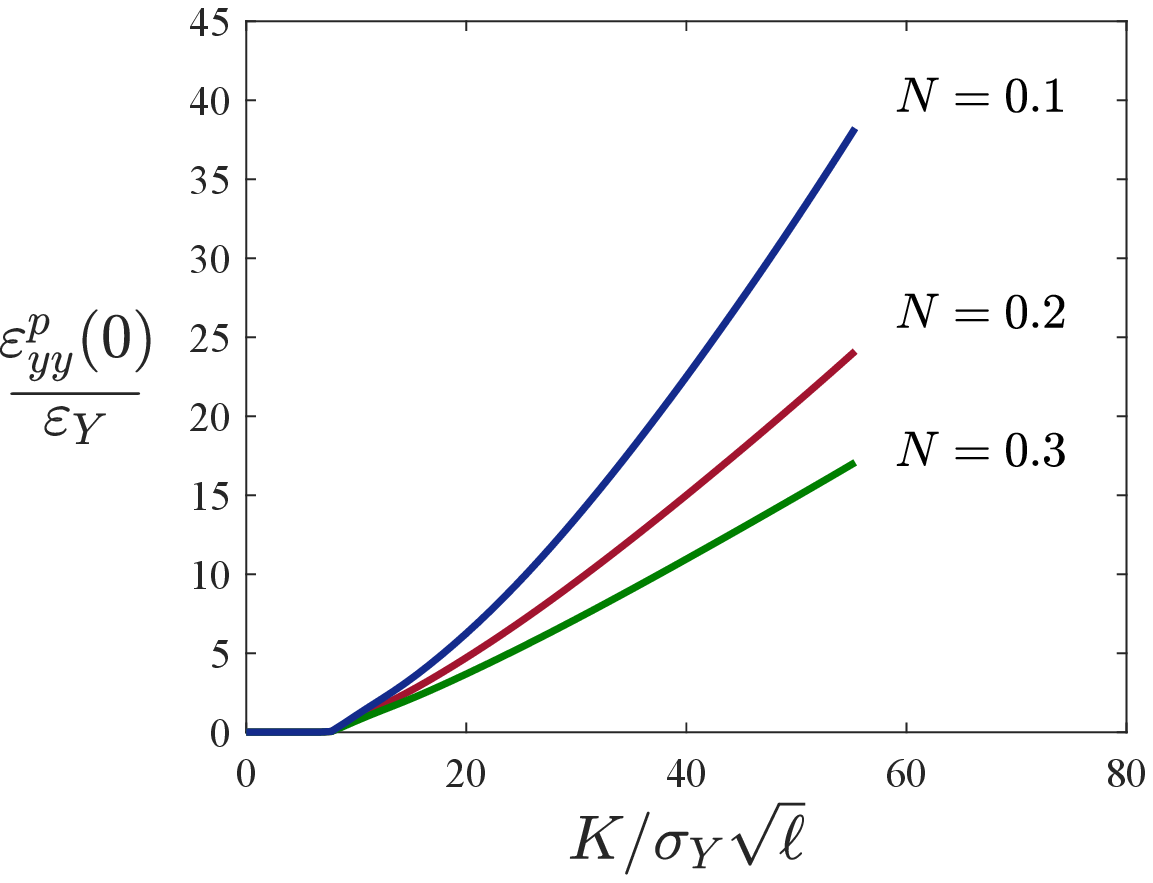}
\caption{Crack tip plastic strain component $\varepsilon_{yy}^p$ as a function of the remote load for different strain hardening exponents. Material properties: $\varepsilon_Y=0.003$, and $\nu=0.3$.}
\label{fig:Ep22App}
\end{figure}

The dependence of $\varepsilon_{yy}^p (r \to 0)$ upon $K/ ( \sigma_Y \sqrt{\ell} )$ is plotted in Fig. \ref{fig:Ep22App} for selected values of $N$. At small $K/ ( \sigma_Y \sqrt{\ell} )$, negligible plasticity exists near the crack tip - the plastic zone vanishes. At larger $K/ ( \sigma_Y \sqrt{\ell} )$ a plastic zone exists and $\varepsilon_{yy}^p (r \to 0)$ increases.

\begin{figure}[H]
\centering
\includegraphics[scale=1.3]{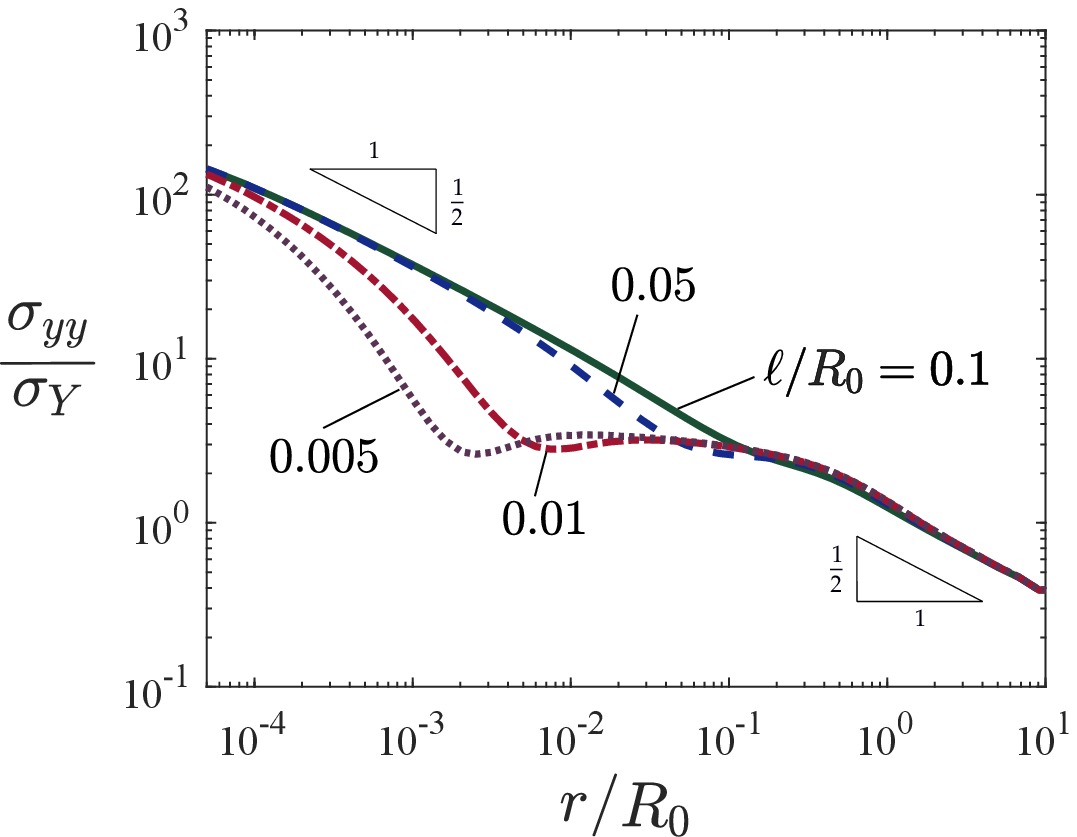}
\caption{Tensile stress ahead of the crack tip for a strain gradient solid with different values of the length scale parameter at $K=20 \sigma_Y \sqrt{\ell}$. Material properties: $N=0.1$, $\varepsilon_Y=0.003$, and $\nu=0.3$.}
\label{fig:S22several}
\end{figure}

The tensile stress component $\sigma_{yy}$ is shown as a function of $r$ in Fig. \ref{fig:S22several} for several values of $\ell/R_0$. The reference size of the plastic zone $R_0$ is given by Irwin's approximation (\ref{Eq:Irwin}). For the strain gradient solid the plastic zone is approximately of size $R_0$ since $\sigma_{yy}/\sigma_Y \approx 1$ at $r/R_0=1$ for all $\ell/R_0$ values considered. Also, the inner elastic zone is of extent $\ell$ to a good approximation. Consequently, the active plastic zone exists between $r \sim \ell$ and $r \sim R_0$.\\

\subsection{Influence on the crack profile and in inhibiting plasticity}

\begin{figure}[H]
\centering
\includegraphics[scale=1.1]{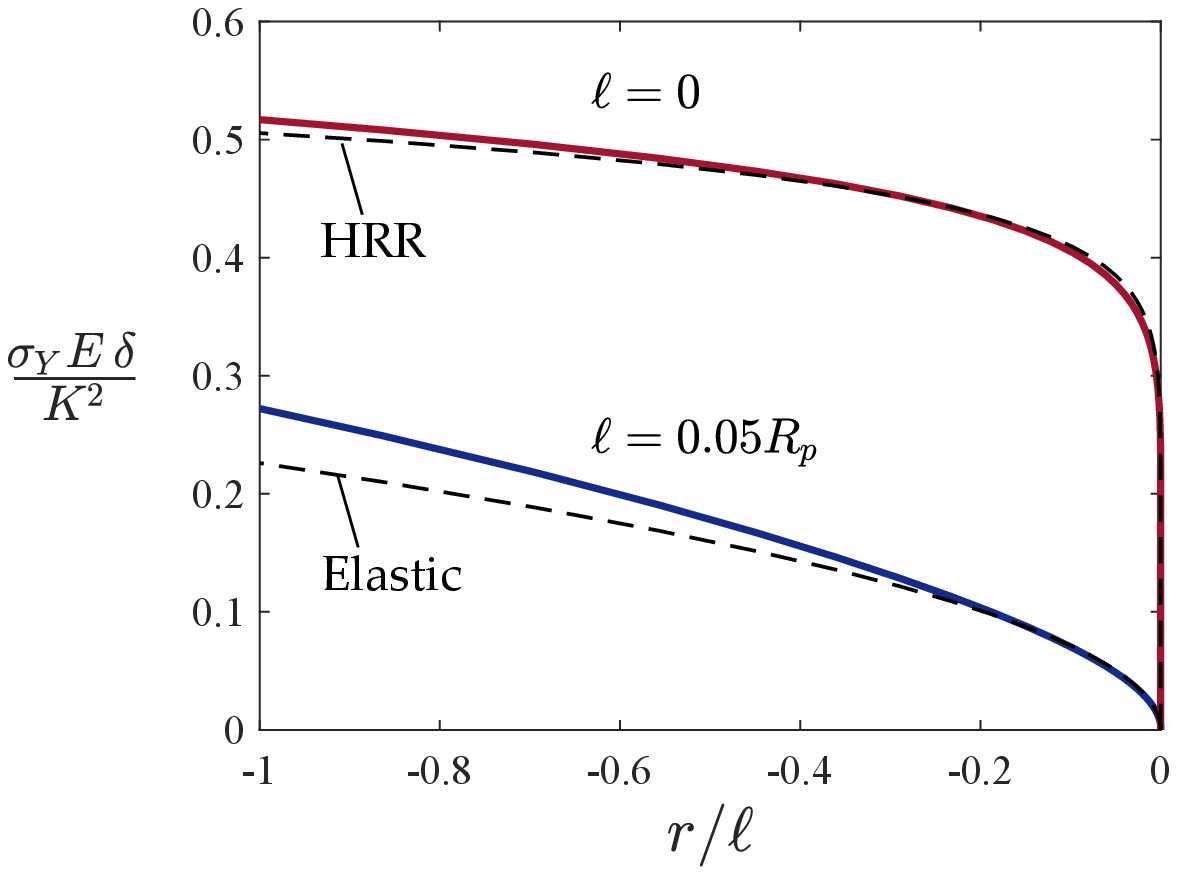}
\caption{Crack opening profile for strain gradient plasticity and conventional plasticity.  Material properties: $N=0.1$, $\varepsilon_Y=0.003$, and $\nu=0.3$.}
\label{fig:CTOD}
\end{figure}

Strain gradient plasticity influences the crack tip profile $\delta (r)$ behind the crack tip. Fig. \ref{fig:CTOD} shows the crack opening profile for conventional ($\ell=0$) and strain gradient plasticity ($\ell=0.05R_0$), along with the solutions from the HRR field and from linear elasticity. The HRR field crack opening profile is given by
\begin{equation}
\frac{ \sigma_Y E \delta}{K^2} = \alpha \left( \frac{r \sigma_Y E}{K^2} \right)^{\frac{N}{N+1}}
\end{equation}

\noindent while the elastic solution reads
\begin{equation}
\frac{ \sigma_Y E \delta}{K^2} = \beta \left( \frac{r \sigma_Y E}{K^2} \right)^{\frac{1}{2}}
\end{equation}

\noindent with $\alpha=0.18$ and $\beta=0.48$. 
The finite element results show large differences between conventional and gradient-enhanced plasticity solutions. Strain gradient plasticity sharpens the crack profile to resemble that of an elastic solid.\\

\begin{figure}[H]
\centering
\includegraphics[scale=1.1]{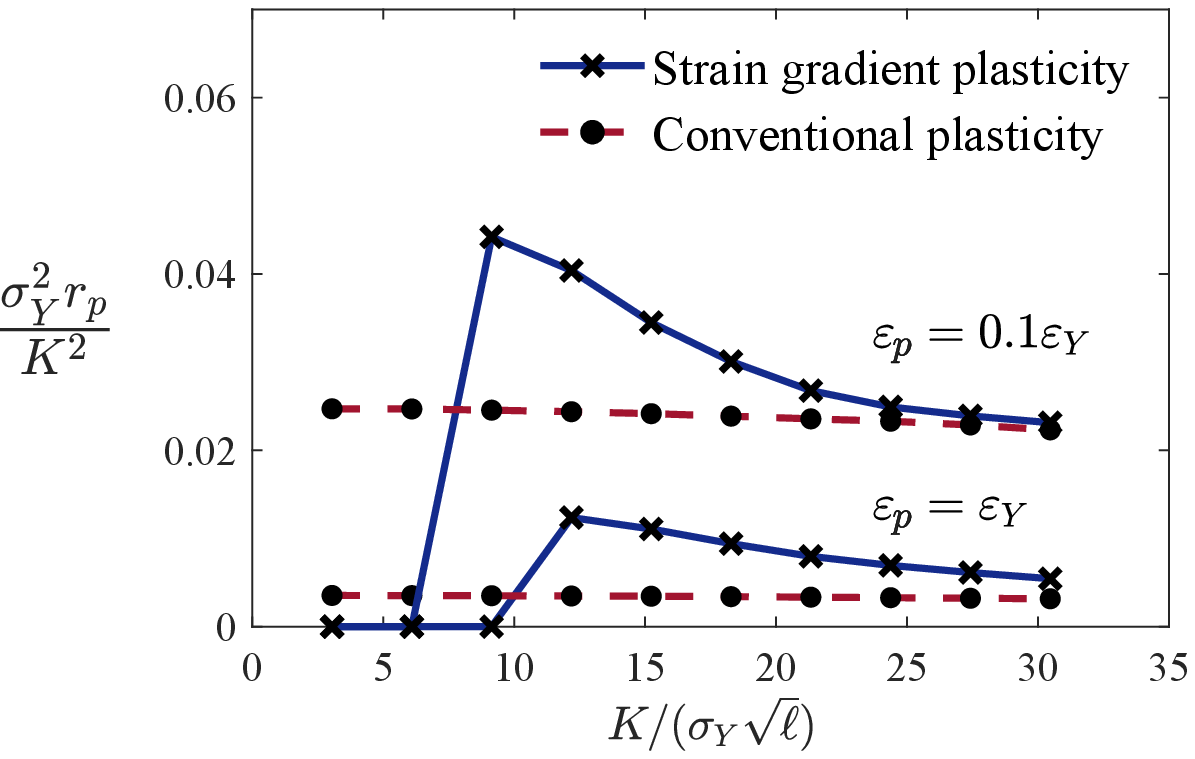}
\caption{Plastic zone size as a function of the remote load for conventional and strain
gradient plasticity. Material properties: $N=0.1$, $\varepsilon_Y=0.003$, and $\nu=0.3$.}
\label{fig:rpVSk}
\end{figure}

Now consider the sensitivity of the plastic zone size $r_p$ to the magnitude of $K/(\sigma_Y \sqrt{\ell})$. We have already noted that, when $K/(\sigma_Y \sqrt{\ell})$ is significantly large, the plastic zone size scales with Irwin's approximation $R_0$ as given by (\ref{Eq:Irwin}). In contrast, when $K/(\sigma_Y \sqrt{\ell})$ is small, we anticipate that the inner elastic core of dimension $\ell$ dominates the plastic zone; this is shown in Fig. \ref{fig:rpVSk}. In order to define the size of the plastic zone $r_p$, a criterion is needed for active yielding. Here, we assume that the plastic zone extends to either the location where $\varepsilon_p/\varepsilon_Y=0.1$ or 1, see Fig. \ref{fig:rpVSk}. It is clear from the figure that the plastic zone size $r_p$ scales with $K^2/\sigma_Y^2$ in the same manner as the conventional elastic-plastic solid for sufficiently large $K/(\sigma_Y \sqrt{\ell})$. However, at small $K/(\sigma_Y \sqrt{\ell})$, on the order of 5 to 10, the plastic zone vanishes. At an intermediate value of $K/(\sigma_Y \sqrt{\ell})$ the active plastic zone for the strain gradient solid is somewhat larger than that predicted for the conventional solid. 

\subsection{Regime of J-dominance}

\begin{figure}[H]
\centering
\includegraphics[scale=1.1]{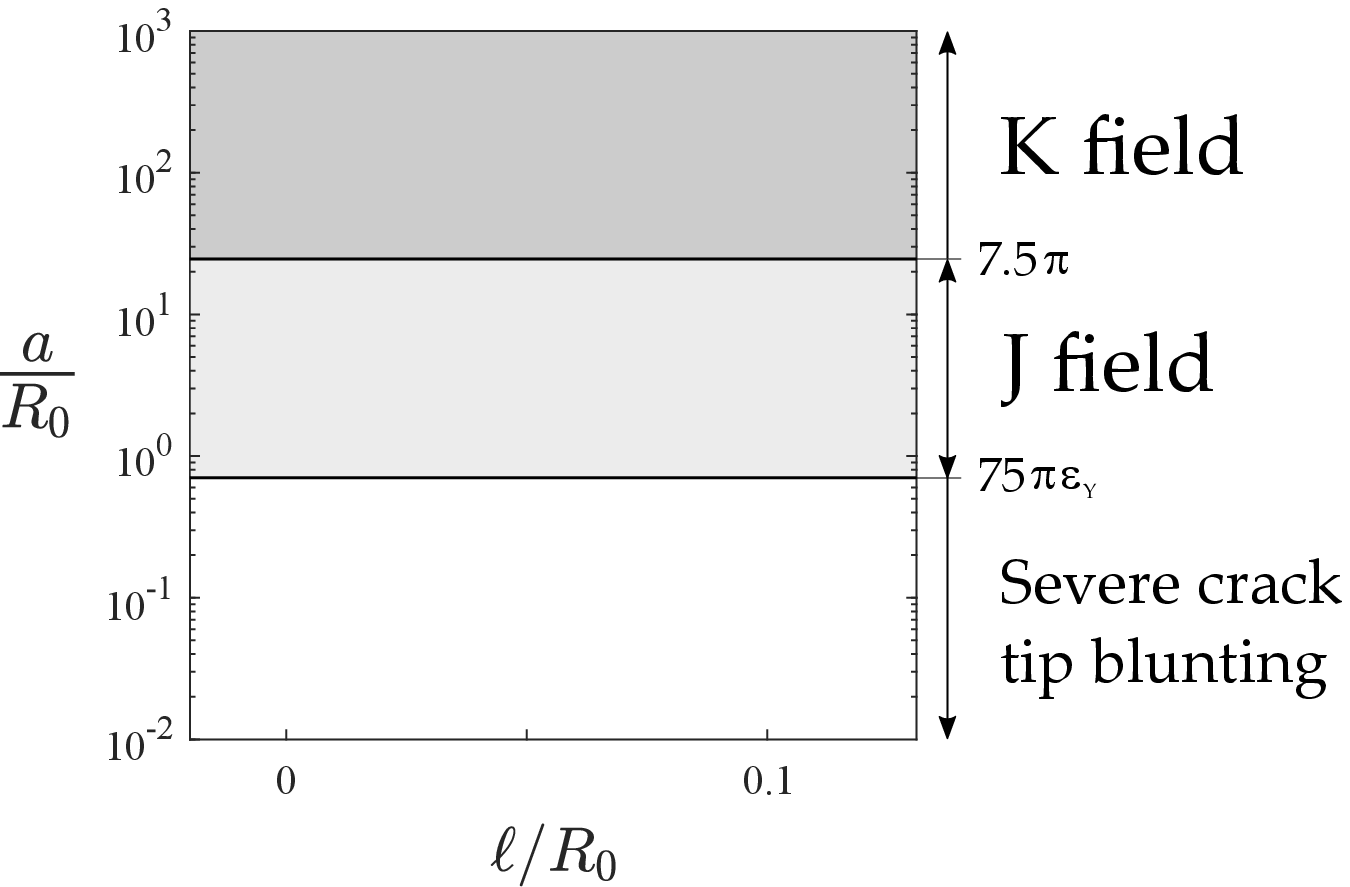}
\caption{Schematic diagram of the regimes and competing length scales involved in the response ahead of a stationary crack.}
\label{fig:SketchLengthScales}
\end{figure}

The small scale yielding (SSY) approach is valid provided the crack length is much greater than the plastic zone size $R_0$ at the onset of fracture, $a > 7.5 \pi R_0$. Thus, on a map with axes $a/R_0$ and $\ell/R_0$ the small scale yielding regime exists for $a/R_0 > 7.5 \pi$; this is shown explicitly in Fig. \ref{fig:SketchLengthScales}. If $a/R_0$ is in the range $75 \pi \varepsilon_Y<a/R_0<7.5 \pi$, then a $J$-field exists near the crack tip and the valid loading parameter becomes $J$ instead of $K$. This regime of $J$-dominance is also sketched in Fig. \ref{fig:SketchLengthScales}. We proceed to explore the stress state near the crack tip for the case of $J$-dominance. To do so we consider a deeply notched beam in three point bending and calculate the tensile stress state ahead of the crack tip.\\

\begin{figure}[H]
\centering
\includegraphics[scale=2.2]{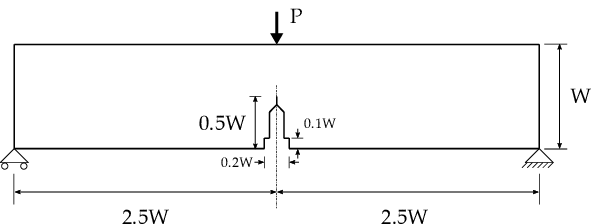}
\caption{Configuration and dimensions of the three point single edge bend specimen.}
\label{fig:SketchBending}
\end{figure}

We follow the ASTM E 1820-01 Standard\footnote{Standard No. ASTM E 1820-01 ``Standard Test Method for Measurement of Fracture Toughness,'' American Society for Testing and Materials, Philadelphia, PA.} and model a three point single edge bend specimen, as outlined in Fig. \ref{fig:SketchBending}. We take advantage of symmetry and model only half of the specimen, with a total of 24000 quadratic quadrilateral plane strain elements. The $J$-integral is computed following the ASTM E 1820 Standard,
\begin{equation}
J=J^e+J^p
\end{equation}

\noindent with $J^e$ being computed from the remote load and the specimen dimensions and $J^p$ being calculated from the area below the force versus displacement curve. A reference length scale $R_0$ can be defined from the estimated value of $J$ as
\begin{equation}\label{Eq:IrwinJ}
R_0 =\frac{1}{3 \pi \left( 1 - \nu^2 \right)} \frac{E J}{\sigma_Y^2}
\end{equation}

\begin{figure}[H]
\centering
\includegraphics[scale=1.1]{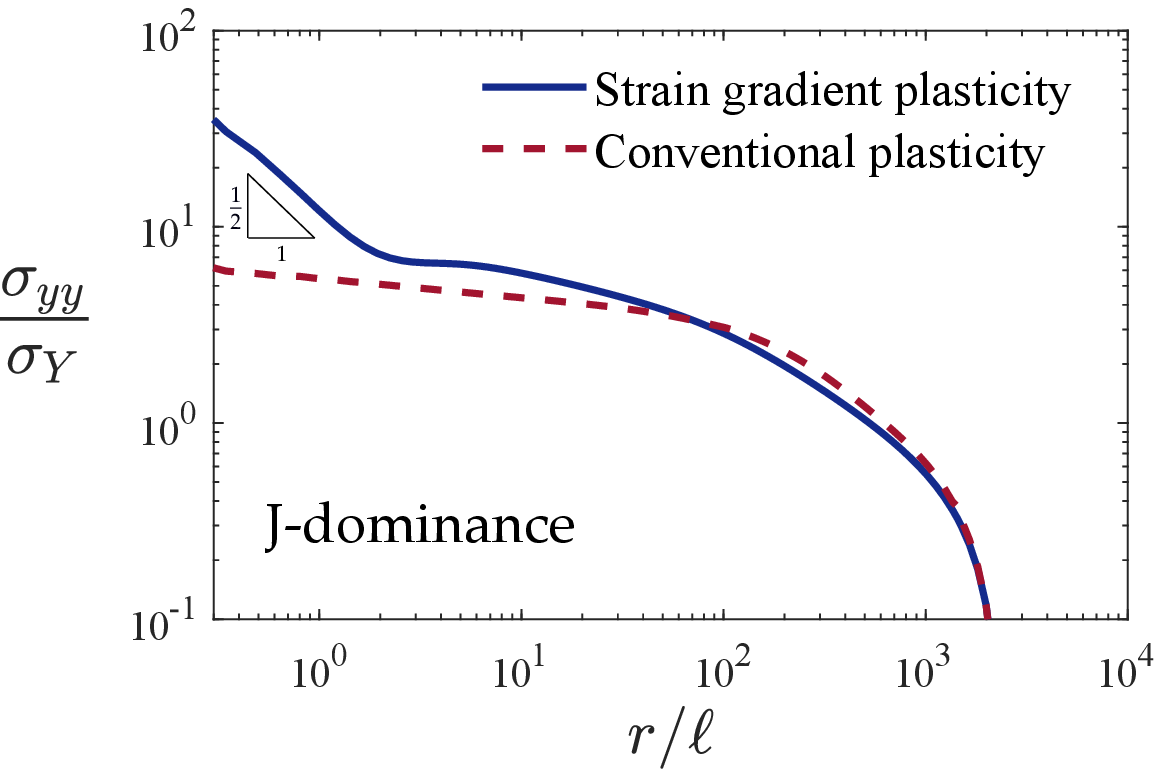}
\caption{Tensile stresses ahead of the crack tip for conventional and strain gradient ($\ell/R_0=0.1$) plasticity under $J$-dominance conditions. Material properties: $N=0.1$, $\varepsilon_Y=0.003$, $\nu=0.3$, and $a/R_0=0.8$.}
\label{fig:JS22vsR}
\end{figure}

Crack tip stresses for $a/R_0=0.8$ and $W/R_0=1.6$ are shown in Fig. \ref{fig:JS22vsR} for strain gradient plasticity ($\ell/R_0=0.1$) and conventional plasticity theory. Finite element results reveal that the elastic core is still present for the case of $J$-dominance; the strain gradient plasticity prediction exhibits the elastic $1/\sqrt{r}$ singularity as $r \to 0$. Thus short cracks, where small scale yielding does not apply, also exhibit an elastic stress state near the crack tip.

\section{Conclusions}
\label{Sec:Concluding remarks}

We examine, numerically and analytically, the crack tip asymptotic response in metallic materials. The solid is characterised by strain gradient plasticity theory, aiming to phenomenologically link scales in fracture mechanics by incorporating the stress elevation due to dislocation hardening. Results reveal that an elastic zone is present in the immediate vicinity of the crack tip. The stresses follow the linear elastic $r^{-1/2}$ singularity and the plastic strains reach a plateau at a distance to the crack tip that scales with the length scale of strain gradient plasticity $\ell$. The dominant role of elastic strains in the vicinity of the crack invalidates asymptotic analyses that neglect their contribution to the total strains. The existence of an elastic core is reminiscent of a dislocation free zone, as introduced by \citet{Suo1993}. \\

The emergence of an elastic core has important implications on the onset of plasticity and the crack opening profile. Numerical predictions show that strain gradient plasticity sharpens the crack opening profile to that of an elastic solid. Differences with conventional plasticity are substantial and results suggest that an experimental characterisation of the crack opening profile could be used to infer the value of the length scale parameter. On the other hand, plasticity is precluded if the remote load is not sufficiently large, such that the plastic zone size (as given by, e.g., Irwin's approximation) falls within the elastic core domain. \\

In addition, we show that the inner elastic regime is also present when the crack is small and an outer elastic $K$ field does not exist. A generalised $J$-integral is presented for strain gradient solids and the stress fields are computed under $J$-dominance conditions in a three point single edge bend specimen.\\

Finally, we note that the material length scale $\ell$ is on the order of a few microns for most metals. This is roughly the smallest scale at which void nucleation and growth occur, suggesting that the transition to an inner zone dominated by elasticity will have important implications in quasi-cleavage but play a secondary role in ductile fracture.
 
\section{Acknowledgements}
\label{Acknowledge of funding}

The authors would like to acknowledge financial support from the European Research Council in the form of an Advance Grant (MULTILAT, 669764). The authors would also like to acknowledge the funding and technical support from BP (ICAM02ex) through the BP International Centre for Advanced Materials (BP-ICAM).



\bibliographystyle{elsarticle-harv}
\bibliography{library}

\begin{thebibliography}{35}
\expandafter\ifx\csname natexlab\endcsname\relax\def\natexlab#1{#1}\fi
\expandafter\ifx\csname url\endcsname\relax
  \def\url#1{\texttt{#1}}\fi
\expandafter\ifx\csname urlprefix\endcsname\relax\def\urlprefix{URL }\fi

\bibitem[{Aifantis(1984)}]{Aifantis1984}
Aifantis, E.~C., 1984. {On the Microstructural Origin of Certain Inelastic
  Models}. Journal of Engineering Materials and Technology 106~(4), 326.

\bibitem[{Ashby(1970)}]{Ashby1970}
Ashby, M.~F., 1970. {The deformation of plastically non-homogeneous materials}.
  Philosophical Magazine 21~(170), 399--424.

\bibitem[{Brinckmann and Siegmund(2008)}]{Brinckmann2008}
Brinckmann, S., Siegmund, T., 2008. {Computations of fatigue crack growth with
  strain gradient plasticity and an irreversible cohesive zone model}.
  Engineering Fracture Mechanics 75~(8), 2276--2294.

\bibitem[{Chen et~al.(1999)Chen, Wei, Huang, Hutchinson, and Hwang}]{Chen1999}
Chen, J.~Y., Wei, Y., Huang, Y., Hutchinson, J.~W., Hwang, K.~C., 1999. {The
  crack tip fields in strain gradient plasticity: the asymptotic and numerical
  analyses}. Engineering Fracture Mechanics 64~(5), 625--648.

\bibitem[{Eshelby(1956)}]{Eshelby1956}
Eshelby, J.~D., 1956. {The Continuum Theory of Lattice Defects}. Solid State
  Physics 3~(C), 79--144.

\bibitem[{Fleck and Hutchinson(2001)}]{Fleck2001}
Fleck, N.~A., Hutchinson, J.~W., 2001. {A reformulation of strain gradient
  plasticity}. Journal of the Mechanics and Physics of Solids 49~(10),
  2245--2271.

\bibitem[{Fleck et~al.(1994)Fleck, Muller, Ashby, and Hutchinson}]{Fleck1994}
Fleck, N.~A., Muller, G.~M., Ashby, M.~F., Hutchinson, J.~W., 1994. {Strain
  gradient plasticity: Theory and Experiment}. Acta Metallurgica et Materialia
  42~(2), 475--487.

\bibitem[{Fleck and Willis(2009)}]{Fleck2009}
Fleck, N.~A., Willis, J.~R., 2009. {A mathematical basis for strain-gradient
  plasticity theory. Part II: Tensorial plastic multiplier}. Journal of the
  Mechanics and Physics of Solids 57~(7), 1045--1057.

\bibitem[{Gao et~al.(1999)Gao, Hang, Nix, and Hutchinson}]{Gao1999}
Gao, H., Hang, Y., Nix, W.~D., Hutchinson, J.~W., 1999. {Mechanism-based strain
  gradient plasticity - I. Theory}. Journal of the Mechanics and Physics of
  Solids 47~(6), 1239--1263.

\bibitem[{Gudmundson(2004)}]{Gudmundson2004}
Gudmundson, P., 2004. {A unified treatment of strain gradient plasticity}.
  Journal of the Mechanics and Physics of Solids 52~(6), 1379--1406.

\bibitem[{Gurtin and Anand(2005)}]{Gurtin2005}
Gurtin, M.~E., Anand, L., 2005. {A theory of strain-gradient plasticity for
  isotropic, plastically irrotational materials. Part I: Small deformations}.
  International Journal of the Mechanics and Physics of Solids 53, 1624--1649.

\bibitem[{Huang et~al.(1997)Huang, Zhang, Guo, and Hwang}]{Huang1997}
Huang, Y., Zhang, L., Guo, T.~F., Hwang, K.~C., 1997. {Mixed mode near-tip
  fields for cracks in materials with strain-gradient effects}. Journal of the
  Mechanics and Physics of Solids 45~(3), 439--465.

\bibitem[{Hutchinson(1968)}]{Hutchinson1968}
Hutchinson, J.~W., 1968. {Singular behaviour at the end of a tensile crack in a
  hardening material}. Journal of the Mechanics and Physics of Solids 16~(1),
  13--31.

\bibitem[{Jiang et~al.(2001)Jiang, Huang, Zhuang, and Hwang}]{Jiang2001}
Jiang, H., Huang, Y., Zhuang, Z., Hwang, K.~C., 2001. {Fracture in
  mechanism-based strain gradient plasticity}. Journal of the Mechanics and
  Physics of Solids 49~(5), 979--993.

\bibitem[{Komaragiri et~al.(2008)Komaragiri, Agnew, Gangloff, and
  Begley}]{Komaragiri2008}
Komaragiri, U., Agnew, S.~R., Gangloff, R.~P., Begley, M.~R., 2008. {The role
  of macroscopic hardening and individual length-scales on crack tip stress
  elevation from phenomenological strain gradient plasticity}. Journal of the
  Mechanics and Physics of Solids 56~(12), 3527--3540.

\bibitem[{Mart{\'{i}}nez-Pa{\~{n}}eda and Beteg{\'{o}}n(2015)}]{IJSS2015}
Mart{\'{i}}nez-Pa{\~{n}}eda, E., Beteg{\'{o}}n, C., 2015. {Modeling damage and
  fracture within strain-gradient plasticity}. International Journal of Solids
  and Structures 59, 208--215.

\bibitem[{Mart{\'{i}}nez-Pa{\~{n}}eda
  et~al.(2017{\natexlab{a}})Mart{\'{i}}nez-Pa{\~{n}}eda, del Busto, and
  Beteg{\'{o}}n}]{TAFM2017}
Mart{\'{i}}nez-Pa{\~{n}}eda, E., del Busto, S., Beteg{\'{o}}n, C.,
  2017{\natexlab{a}}. {Non-local plasticity effects on notch fracture
  mechanics}. Theoretical and Applied Fracture Mechanics 92, 276--287.

\bibitem[{Mart{\'{i}}nez-Pa{\~{n}}eda
  et~al.(2016{\natexlab{a}})Mart{\'{i}}nez-Pa{\~{n}}eda, del Busto, Niordson,
  and Beteg{\'{o}}n}]{IJHE2016}
Mart{\'{i}}nez-Pa{\~{n}}eda, E., del Busto, S., Niordson, C.~F., Beteg{\'{o}}n,
  C., 2016{\natexlab{a}}. {Strain gradient plasticity modeling of hydrogen
  diffusion to the crack tip}. International Journal of Hydrogen Energy
  41~(24), 10265--10274.

\bibitem[{Mart{\'{i}}nez-Pa{\~{n}}eda et~al.(2019)Mart{\'{i}}nez-Pa{\~{n}}eda,
  Deshpande, Niordson, and Fleck}]{JMPS2019}
Mart{\'{i}}nez-Pa{\~{n}}eda, E., Deshpande, V.~S., Niordson, C.~F., Fleck,
  N.~A., 2019. {The role of plastic strain gradients in the crack growth
  resistance of metals}. Journal of the Mechanics and Physics of Solids (in
  press).

\bibitem[{Mart{\'{i}}nez-Pa{\~{n}}eda
  et~al.(2017{\natexlab{b}})Mart{\'{i}}nez-Pa{\~{n}}eda, Natarajan, and
  Bordas}]{CM2017}
Mart{\'{i}}nez-Pa{\~{n}}eda, E., Natarajan, S., Bordas, S., 2017{\natexlab{b}}.
  {Gradient plasticity crack tip characterization by means of the extended
  finite element method}. Computational Mechanics 59, 831--842.

\bibitem[{Mart{\'{i}}nez-Pa{\~{n}}eda and Niordson(2016)}]{IJP2016}
Mart{\'{i}}nez-Pa{\~{n}}eda, E., Niordson, C.~F., 2016. {On fracture in finite
  strain gradient plasticity}. International Journal of Plasticity 80,
  154--167.

\bibitem[{Mart{\'{i}}nez-Pa{\~{n}}eda
  et~al.(2016{\natexlab{b}})Mart{\'{i}}nez-Pa{\~{n}}eda, Niordson, and
  Gangloff}]{AM2016}
Mart{\'{i}}nez-Pa{\~{n}}eda, E., Niordson, C.~F., Gangloff, R.~P.,
  2016{\natexlab{b}}. {Strain gradient plasticity-based modeling of hydrogen
  environment assisted cracking}. Acta Materialia 117, 321--332.

\bibitem[{Nielsen et~al.(2012)Nielsen, Niordson, and Hutchinson}]{Nielsen2012}
Nielsen, K.~L., Niordson, C.~F., Hutchinson, J.~W., 2012. {Strain gradient
  effects on steady state crack growth in rate-sensitive materials}.
  Engineering Fracture Mechanics 96, 61--71.

\bibitem[{Nix and Gao(1998)}]{Nix1998}
Nix, W.~D., Gao, H.~J., 1998. {Indentation size effects in crystalline
  materials: A law for strain gradient plasticity}. Journal of the Mechanics
  and Physics of Solids 46~(3), 411--425.

\bibitem[{Panteghini and Bardella(2016)}]{Panteghini2016}
Panteghini, A., Bardella, L., 2016. {On the Finite Element implementation of
  higher-order gradient plasticity, with focus on theories based on plastic
  distortion incompatibility}. Computer Methods in Applied Mechanics and
  Engineering 310, 840--865.

\bibitem[{Poole et~al.(1996)Poole, Ashby, and Fleck}]{Poole1996}
Poole, W.~J., Ashby, M.~F., Fleck, N.~A., 1996. {Micro-hardness of annealed and
  work-hardened copper polycrystals}. Scripta Materialia 34~(4), 559--564.

\bibitem[{Pribe et~al.(2019)Pribe, Siegmund, Tomar, and Kruzic}]{Pribe2019}
Pribe, J.~D., Siegmund, T., Tomar, V., Kruzic, J.~J., 2019. {Plastic strain
  gradients and transient fatigue crack growth: a computational study}.
  International Journal of Fatigue 120, 283--293.

\bibitem[{Rice(1968)}]{Rice1968a}
Rice, J.~R., 1968. {Mathematical Analysis in the Mechanics of Fracture}.
  Mathematical Fundamentals 2~(B2), 191--311.

\bibitem[{Rice and Rosengren(1968)}]{Rice1968}
Rice, J.~R., Rosengren, G.~F., 1968. {Plane strain deformation near a crack tip
  in a power-law hardening material}. Journal of the Mechanics and Physics of
  Solids 16~(1), 1--12.

\bibitem[{St{\"{o}}lken and Evans(1998)}]{Stolken1998}
St{\"{o}}lken, J.~S., Evans, A.~G., 1998. {A microbend test method for
  measuring the plasticity length scale}. Acta Materialia 46~(14), 5109--5115.

\bibitem[{Suo et~al.(1993)Suo, Shih, and Varias}]{Suo1993}
Suo, Z., Shih, C.~F., Varias, A.~G., 1993. {A theory for cleavage cracking in
  the presence of plastic flow}. Acta Metallurgica Et Materialia 41~(5),
  1551--1557.

\bibitem[{Tvergaard and Niordson(2008)}]{Tvergaard2008}
Tvergaard, V., Niordson, C.~F., 2008. {Size effects at a crack-tip interacting
  with a number of voids}. Philosophical Magazine 88~(30-32), 3827--3840.

\bibitem[{Wei and Hutchinson(1997)}]{Wei1997}
Wei, Y., Hutchinson, J.~W., 1997. {Steady-state crack growth and work of
  fracture for solids characterized by strain gradient plasticity}. Journal of
  the Mechanics and Physics of Solids 45~(8), 1253--1273.

\bibitem[{Williams(1957)}]{Williams1957}
Williams, M.~L., 1957. {On the stress distribution at the base of a stationary
  crack}. Journal of Applied Mechanics 24, 109--114.

\bibitem[{Xia and Hutchinson(1996)}]{Xia1996}
Xia, Z.~C., Hutchinson, J.~W., 1996. {Crack tip fields in strain gradient
  plasticity}. Journal of the Mechanics and Physics of Solids 44~(10),
  1621--1648.

\end{thebibliography}


\end{document}